\documentclass[english,aps,preprint]{revtex4}
\usepackage[T1]{fontenc}
\usepackage[latin9]{inputenc}
\usepackage{amsmath}
\usepackage{graphicx}
\usepackage{esint}

\makeatletter

\DeclareFontEncoding{T3}{}{}
\DeclareSymbolFont{tipasymb}{T3}{cmr}{m}{n}

\DeclareMathAccent{\dgrave}{\mathord}{tipasymb}{'15}

\@ifundefined{textcolor}{}
{%
 \definecolor{BLACK}{gray}{0}
 \definecolor{WHITE}{gray}{1}
 \definecolor{RED}{rgb}{1,0,0}
 \definecolor{GREEN}{rgb}{0,1,0}
 \definecolor{BLUE}{rgb}{0,0,1}
 \definecolor{CYAN}{cmyk}{1,0,0,0}
 \definecolor{MAGENTA}{cmyk}{0,1,0,0}
 \definecolor{YELLOW}{cmyk}{0,0,1,0}
}

\makeatother

\usepackage{babel}
\begin{document}

\title{A Chaotic, Deterministic Model for Quantum Mechanics}

\author{Carlton Frederick}

\email{carl@frithrik.com}

\homepage{http://www.frithrik.com}

\affiliation{Central Research Group, 127 Pine Tree Road, Ithaca, NY 14850 }
\begin{abstract}
With the decline of the Copenhagen interpretation of quantum mechanics
and the recent experiments indicating that quantum mechanics \emph{does}
actually embody 'objective reality', one might ask if a 'mechanical',
conceptual model for quantum mechanics could be found. We propose
such a model.

A previous paper \cite{F1} noted that space-time vacuum energy fluctuations
implied mass fluctuations and, through general relativity, curvature
fluctuations. And those fluctuations are indicated by fluctuations
of the metric tensor. The metric tensor fluctuations, there presumed
to be described by stochastic variables in the tensor elements, can
'explain' the uncertainty relations and non-commuting properties of
conjugate variables. It also argues that the probability density $\Psi^{*}\Psi$
is proportional to the square root of minus the determinant of the
metric tensor (the differential volume element) $\sqrt{-\left\Vert g_{_{\mu\nu}}\right\Vert }$.

The present paper extends those ideas by arguing that the metric elements
are actually not stochastic but are oscillating at a sufficiently
high frequency that measured values of same appear stochastic (i.e.
crypto-stochastic). This is required to allow that the position probability
density $\sqrt{-\left\Vert g_{_{\mu\nu}}\right\Vert }$ be a non-stochastic
variable.

We'll defer the discussion of whether the fluctuations are truly random
or just apparently random, but note that the current best description
of space-time is given by the general relativity field equations.
They are nonlinear and (as they do not describe probabilities) deterministic.
These two features are necessary for chaotic behavour.

We posit that the oscillations at the position of particles are described
as torsional vibrations. A crypto-stochastic (or chaotic) oscillating
metric yields, among other things, a model of superposition, photon
polarization, and entanglement, and all within the confines of a 4-dimensional
space-time. Further, this implies the deBroglie view (as opposed to
the Copenhagen interpretation) that the particle and wave are different
entities. The proposed model is one of 'objective reality' but, of
course, as required by Bell's theorem, at the expense of temporal
locality.
\end{abstract}
\maketitle

\section{introduction}

'Weak measurement' experiments \cite{key-6,key-7,key-5}, building
on the pioneering work of Yakir. Aharonov and Lev Vaidman \cite{key-8},
have dealt a serious if not mortal blow to the Copenhagen interpretation
of quantum mechanics, and has given new life to the DeBroglie-Bohm
Pilot Wave idea, and a re-emergence of objective reality in quantum
mechanics\cite{key-2,key-3,key-4} {[}Abandonment of objective-reality
says that the physical situation is established by measurement (e.g.
the cat is both alive and dead until measured).{]} In the Pilot wave
theory however, the nature and source of those waves has not been
explained. The model we propose attempts to explain pilot waves and
also develop a way of interpreting a wide range of quantum phenomena.
It covers a lot of territory but, by necessity, far from thoroughly.
The idea is to build a conceptual scaffolding for a full theory.

At the most primitive level, we regard quantum mechanics as emergent
from vacuum energy fluctuations throughout space-time, and also from
the paths particles take in the space-time. Since vacuum energy fluctuations
imply (via general relativity) mass fluctuations. Mass produces curvature
which is reflected in fluctuations of the metric tensor components.
To first approximation, we take the metric fluctuations to be stochastic
or chaotic. Spaces with stochastic metrics have been investigated
by Schweizer \cite{F2-2} for metric spaces and March \cite{F4},
\cite{F5} for Minkowski space. Blokhintsev \cite{F6-1} has considered
the physics of a space-time with a small stochastic component. 

The following three parts represent a progression of the model. Deductions
of quantum mechanics structure are explored in each part.

Our attempt in Part I is to show that some of the fundamentals of
quantum mechanics can be deduced by both imposing apparent stochasticity
on the metric tensor and also assuming, and attempting to justify,
a few theoretical postulates. In Part II, we argue that the metric
tensor components \emph{cannot }be stochastic, but only \emph{seem}
stochastic and are actually oscillations at an immeasurably high frequency.
Yet more quantum effects can thereby be explained, including polarization
and entanglement. Indeed, Masreliez \cite{F6-2} has derived the Schrödinger
equation by imposing a type of oscillation on the metric. And in Part
III, we briefly discuss the chaos interpretation.

\part{Stochastic Space-time}

\section{Postulates}

\noindent \emph{1: A Generalization of Mach's Principle}

\noindent 1.1 In the absence of mass, space becomes not flat, but
stochastic (or chaotic).

This is the prime reason for uncertainty in quantum mechanics in this
model, and the (apparent) stochasticity directs particle trajectories,
the probability of which is given by the wave function, $\Psi$. In
Part III, we'll argue that space-time is not stochastic, but chaotic.
The first order effects would be the same, but the philosophy is very
different. Chaos is not stochastic but indeterminate yet deterministic.

\noindent 1.2 The (apparent) stochasticity is manifested by fluctuations
of the metric tensor.

\noindent 1.3 The mass distribution determines not only the space-time
geometry, but also the space-time (apparent) stochasticity.

This and 1.4 encapsulates Mach's Principle.

\noindent 1.4 The more mass in the space-time, the less the space-time
appears stochastic.

\noindent 1.5 At the position of a mass, the space-time does not fluctuate.

We posit this so that masses aren't pulled apart by the metric fluctuations.

\noindent \emph{2: The Contravariant nature of Measurements}

\noindent 2.1 All measurements of dynamical variables correspond to
\emph{contravariant} components of tensors.

We'll attempt to justify this in Section III

\noindent \emph{3: The Probability Density P(x,t) Identification at
a Venue.}

\emph{(Note: we use the term space-time 'venue' instead of space-time
event {[}a point in x,y,z,t{]} to indicate that space-time is 'grainy',
i.e. there is a minimum length and time interval. We assume this to
avoid having the infinite vacuum energy fluctuation that would be
predicted at a point.)}

\noindent 3.1 P(x,t) is proportional to $\sqrt{-\left\Vert g_{\mu\nu}\right\Vert }$,
i.e. the square root of minus the determinant of the metric tensor.

In differential geometry, the quantity $\sqrt{-\left\Vert g_{\mu\nu}\right\Vert }dx^{1}dx^{2}dx^{3}dx^{4}$
corresponds to the Euclidean differential volume element $-dx*dy*dz*dt$.
For a particle traveling space-time, we assume then, that the probability
of the particle being in a particular differential volume element
is proportional to the relative 'size' of the volume element. Note
that here we assume the deBroglie idea that the particle actually
\emph{is} in a particular location and the wave function is (as deBroglie
puts it) the 'ghost wave' that guides the particle. 

There is a major constraint on the model: While the metric tensor
elements have an apparent stochastic component, if it is to be associated
with a probability density, the determinant of the metric tensor,
while allowed to change with time, must not seem stochastic. That
is to say that the probability density is a well-defined (deterministic)
quantity in quantum mechanics. This constraint is addressed in Parts
II and III.

\noindent \emph{4: The Wave Function $\Psi$ Identification at a Venue}

\noindent 4.1 There exists a local complex coordinate system where
the metric tensor is (at a given venue) diagonal and a component of
the metric is the wave function $\Psi$. 

This isn't central to the model but exists simply as an expression
of the idea that at present, there are two separate concepts: the
metric $g_{\mu\nu},$ and the wave function, $\Psi$. It is an aim
of our geometrical approach to express one of these quantities in
terms of the other. We'll address this in Section V.

Incidentally, we also suggest that there is one concept that should
be two: waves from the wave equation. One concept is the wave as an
indicator of probability, and the other concept as the wave identified
with the momentum of the particle. E.g. a wavelength of light and
the wave indicating the probability of the photon being at a particular
location are two different things. We'll (partially) address this
in Part II. 

\noindent \emph{5: Metric Superposition}

\noindent 5.1 If at he position of a particle, the metric due to a
specific physical situation is $g_{\mu\nu}(1)$ and the metric due
to a different physical situation is $g_{\mu\nu}(2)$, then the metric
due to both of the physical situations is $g_{\mu\nu}(3)=\frac{1}{2}\left[g_{\mu\nu}(1)+g_{\mu\nu}(2)\right]$.

This is \emph{linear} superposition but for general relativity, this
is clearly false as the field equations are nonlinear in terms of
the metric tensor. But particle masses are very small (in general
relativity terms) and the particle velocities we consider (where the
mass is greater than zero) are low compared to the speed of light.
So we feel justified in using the \emph{linearized} general relativity
field equations (especially as the electro-weak force is some $10^{39}$
stronger than the gravitational force). The superposition postulate
then, is only an approximation, albeit a very good approximation as,
for quantum mechanical masses, the linearized field equations diverge
only very slightly from the full field equations. 

We regard conventional quantum theory as an extension of Newtonian
mechanics, and hence it is a linear theory. Our model, as it is an
extension of relativity theory, implies a non-linear theory and we
would expect superposition to break down at very high particle energies.

\section{The Contravariant Nature of Measurements}

The contention is that whenever a measurement can be reduced to a
displacement in a coordinate system, it will be represented by contravariant
components in the coordinate system. Of course, if the metric tensor
$g_{\mu\nu}$ is known, one can calculate covariant quantities from
their contravariant counterparts. In our model though, the quantum
fluctuations in the vacuum energy is reflected in apparently stochastic
components of the metric elements. So, if we try to use the metric
tensor to lower the index of a contravariant quantity, the cooresponding
covariant quantity will appear at least partially stochastic.

We will attempt to show that, at least for Minkowski space, measurements
are contravariant. We'll do that by considering an idealized measurement.
Before we do, however, consider as an example the case of measuring
the distance to a Schwartzschild singularity (i.e. a black hole) in
the Galaxy. Let the astronomical distance to the object be $\dgrave{r}$
\ensuremath{\equiv} ($\overline{\xi}^{1}$). The covariant equivalent
of the radial coordinate $r$ is $\xi_{1}$, and

\[
\xi_{1}=g_{1\nu}\xi^{\nu}=g_{11}\xi^{1}=\frac{r}{1-2Gm/r}
\]

so that the contravariant distance to the object is 

\[
distance=\int_{0}^{\overline{r}}d\overline{r}=\overline{r}
\]

whereas the covariant distance is

\[
\overline{\xi}=\int_{0}^{\overline{r}}d\left(\frac{r}{1-2Gm/r}\right)=\infty
\]

From this, it is clear that only the contravariant distance is measurable.

Now as to the postulate, first consider figure 1 showing vector components
in a flat space with an oblique 2-dimensional coordinate system. The
contravariant coordinates of a point V are given by the parallelogram
law of vector addition, while the covariant components are obtained
by orthogonal projection onto the axes \cite{F6}.

\begin{figure}
\includegraphics[scale=0.25]{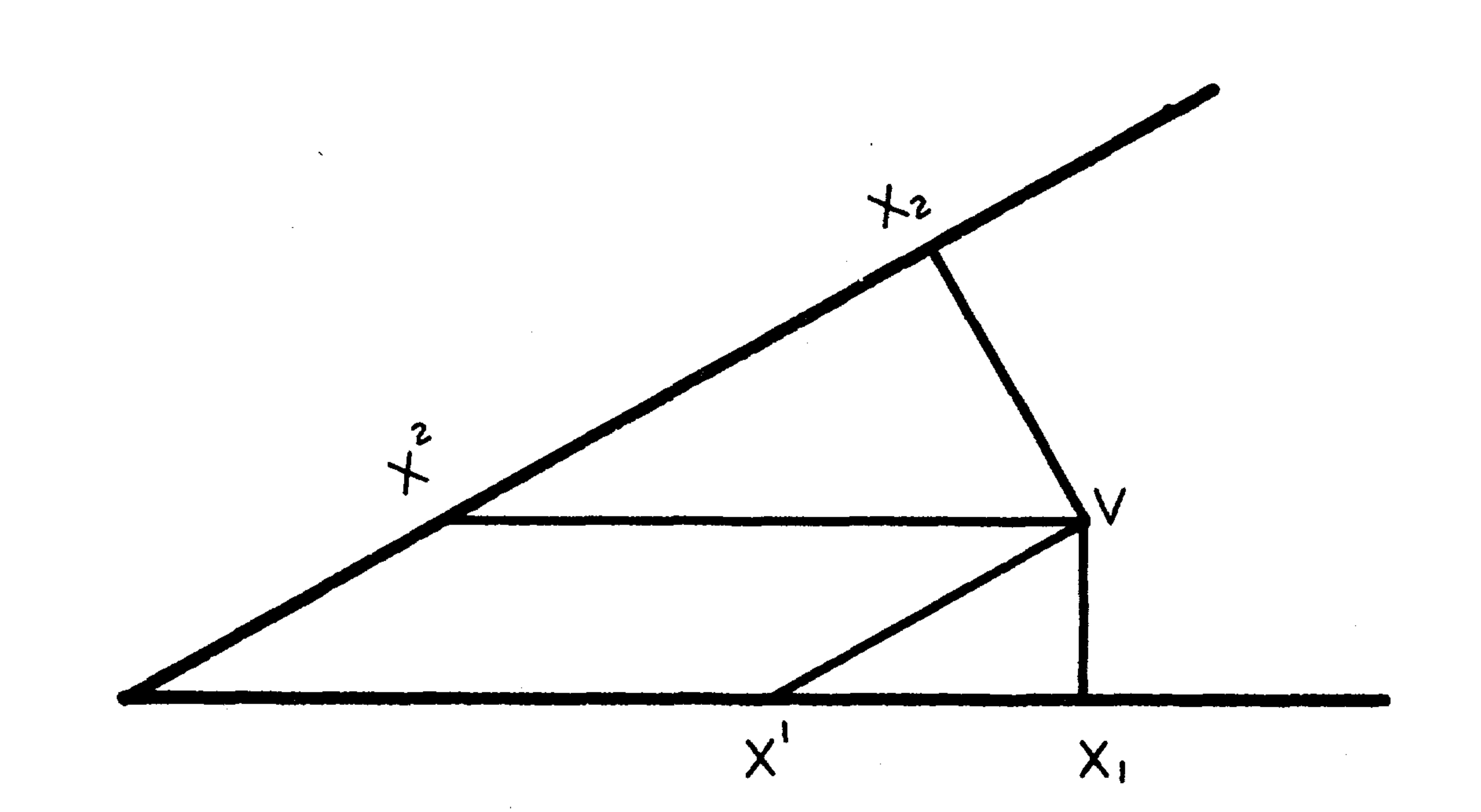}

Fig. 1. Covariant and contravariant components in oblique coordinates
\end{figure}

We shall now consider an idealized measurement in special relativity,
i.e., Minkowski space. Consider the space-time diagram of Fig. 2.
We are given that in the coordinate system $x',t'$, an object (the
line $m.n$) is at rest.

\begin{figure}
\includegraphics[scale=0.3]{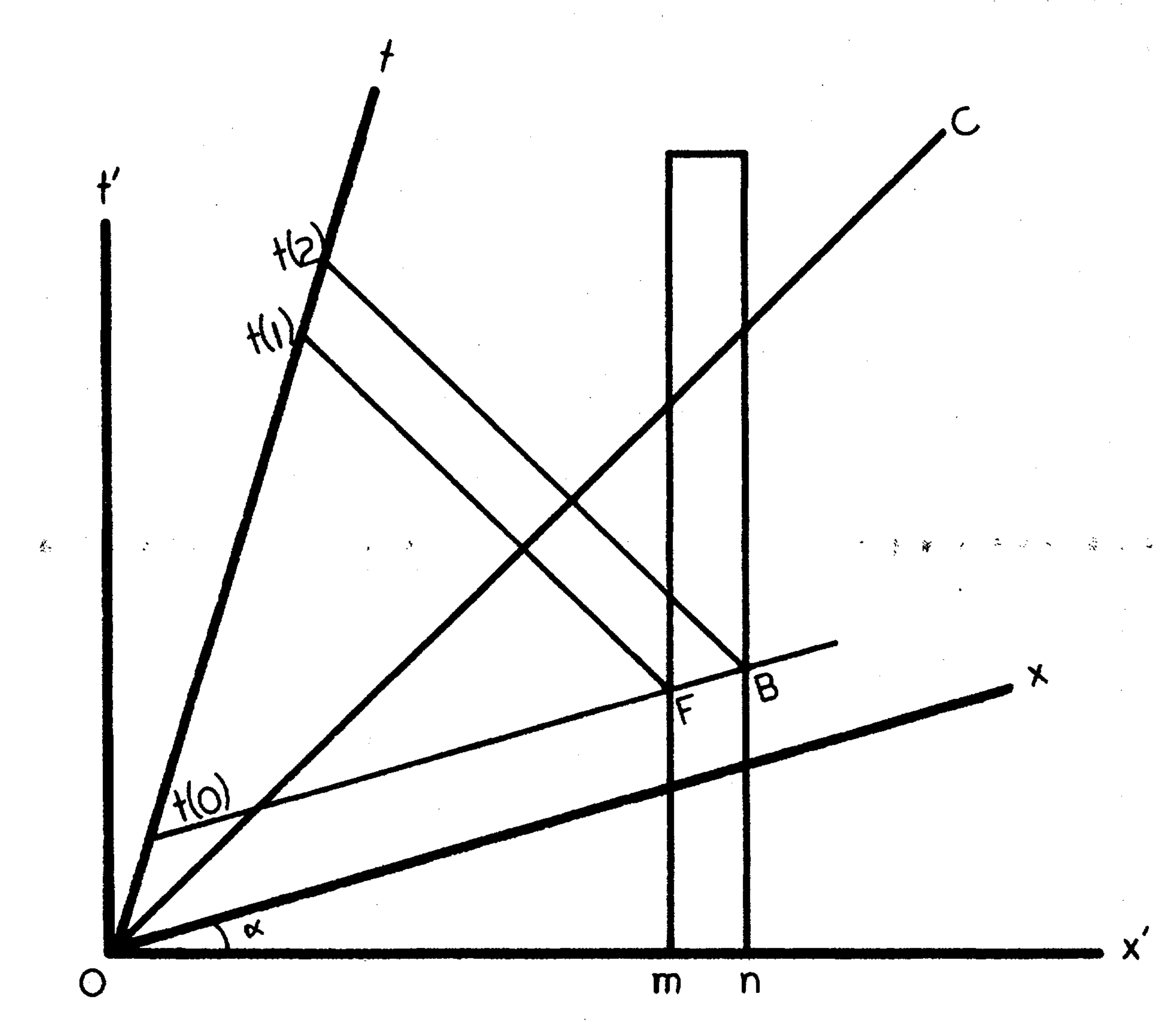}

Fig. 2. An idealized measurement.
\end{figure}

If one considers the situation from a coordinate system $x,t$ traveling
with velocity$v$ along the $x'$axis, one has the usual Minkowski
diagram\cite{F7}with coordinate axes $Ox$and $Ot$ and velocity
$v=tan$ $\alpha$ (where the units are chosen such that the speed
of light is unity). $OC$ is part of the light cone.

Noting that the unprimed system is a suitable coordinate system in
which to work, we now drop from consideration the original $x',t'$
coordinates.

We wish to determine the 'length' of the object in the $x,t$ coordinate
system. At time $t(0)$, let a photon be emitted from each end of
the object (i.e., from points $F$ and $B$). The emitted photons
will intercept the $t$ axis at times $t(1)$ and $t(2)$. We then
can then deduce that the length of the object is $t(2)-t(1)$ (where
c=1). The question is: What increment on the $x$ axis corresponds
to the time interval $t(2)-t(1)$?

Note that the arrangement that the photons be emitted at time $t(0)$
is nontrivial, but that it can be done in principle. For the present,
let us simply assume that there is a person on the object who knows
special relativity and knows how fast the object is moving with respect
to the coordinate system. This person then calculates when to emit
the photons so that they will be emitted simultaneously with respect
to the $x,t$ coordinate system.

Consider now Fig. 3, representing the an analysis of the measurement.

\begin{figure}
\includegraphics[scale=0.3]{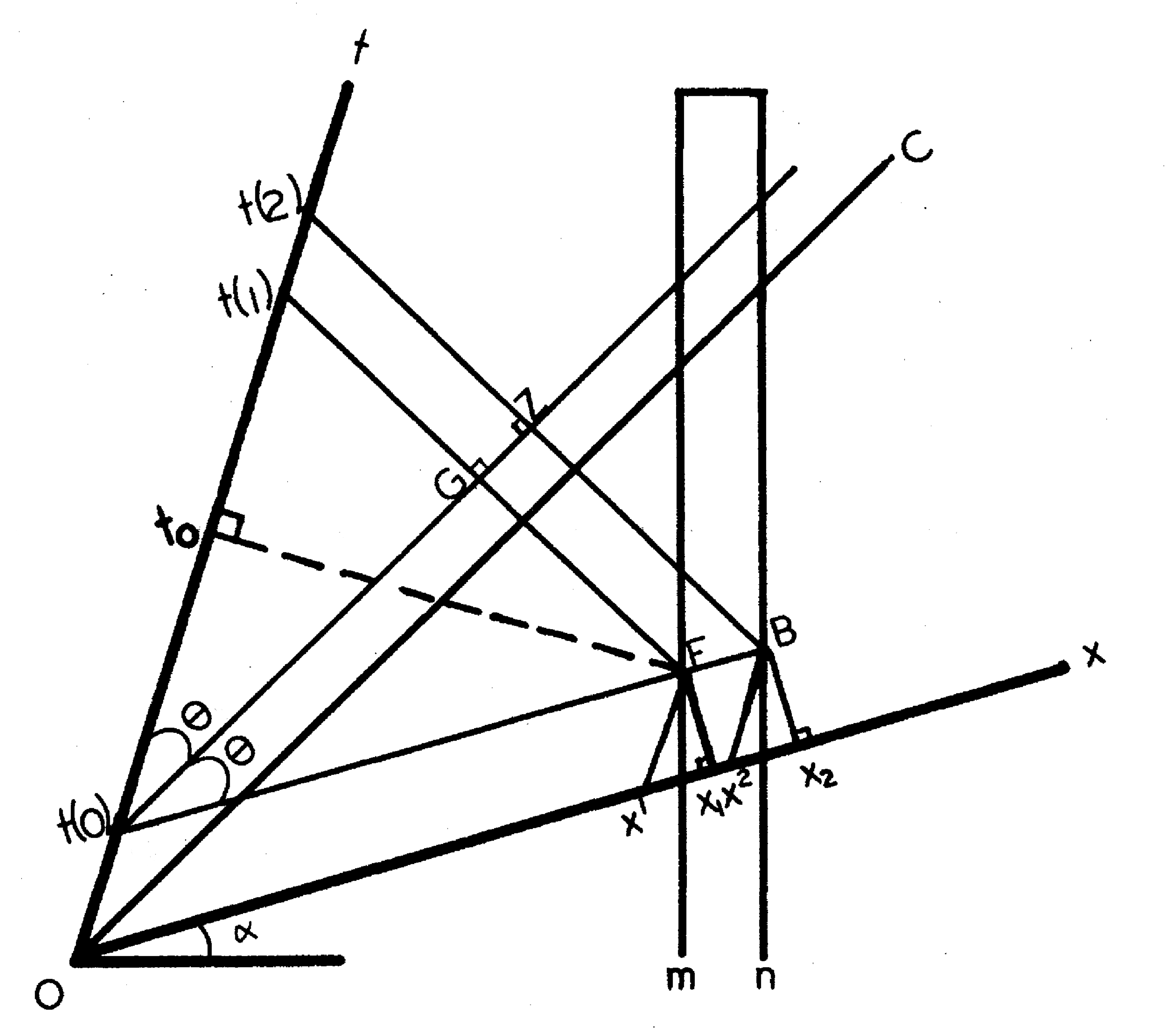}

Fig. 3. Analysis of the idealized measurement.
\end{figure}

The figure is Fig.2 with the addition of the contravariant coordinates
of $F$ and $B$, $x^{1}$and $x^{2}$respectively. It is easily shown
that $t(2)-t(1)=x^{2}-x^{1}$. This is seen by noticing that $x^{2}-x^{1}$
equals the line segment $B,F$, and that triangle $t(2)$, $t(0)$,
$Z$ is congruent to triangle $B$, $t(0)$, $Z$. If we consider
the covariant components, we notice that $x_{2}-x_{1}=x^{2}-x^{1}$.
This is not surprising since coordinate differences (such as $x_{2}-x_{1}$)
behave, in flat space, as contravariant objects\cite{F8}. To address
our postulate, we must consider, not coordinate differences which
automatically satisfy the conjecture, but the coordinates individually.
Consider in Fig 3. a measurement not of the length of the object,
but the position of the trailing edge $m$ of the object $m,n$. Assume
again that at time $t(0)$, a photon is emitted at $F$ and is received
at $t(1)$. The observer could then determine the position of $m$
at $t(0)$ by simply measuring the distance $t(1)-t(0)$ on the $x$
axis. Notice that this is the same as the contravariant coordinate
value $x^{1}$. To determine the corresponding covariant value, here
indicated as $t_{0}$, one would need to know the angle $\alpha$
(which is determined by the metric tensor).

The metric tensor $g_{ik}$ is defined as $\hat{e_{i}}$· $\hat{e_{k}}$
where $\hat{e_{i}}$ and $\hat{e_{k}}$ are the unit vectors in the
directions of the coordinate axes $x^{i}$ and $x^{k}$. Therefore
in order to consider an uncertain metric (in this 2-space), we can
simply consider that the angle $\alpha$ is uncertain. In this case,
measurement $x^{1}$ is still well-defined ($x^{1}=t(1)-t(0)$ ),
but there is no way to determine $x_{1}$ because it is a function
of the angle $\alpha$. In this case then, only the contravariant
components of position are measurable. It is easy to see from the
geometry, that if one were to use the covariant representation of
$t(0)$, $t_{0}$, one could not obtain a metric-free position measure
of $m$.

The above, of course, can't be considered a rigorous proof of the
conjecture that dynamical measurements are only of contravariant quantities,
but it is, I believe, strongly suggestive.

\section{Basic Physics Results from part I}

We derive first the motion of a test particle in the space-time far
from other masses. Overall, the space-time is assumed to have enough
mass to make the space-time, in the large and on the average, Minkowskian.
The requirement that the test particle be far from other masses is
so that we can consider that the space time points (venues) in the
region can be considered indistinguishable.

Consider a space-time particular venue $\Theta_{1}$. Let the metric
tensor at $\Theta_{1}$ be $\tilde{g}_{\mu\nu}$ (a tilda over a symbol
indicates that it is apparently stochastic). Since $\tilde{g}_{\mu\nu}$
seems stochastic, the metric components, by definition, don't have
predictable values. So we cannot know $\tilde{g}_{\mu\nu}$ but we
can ask for $P\left(g_{\mu\nu}\right)$ which is the probability of
a particular metric $g_{\mu\nu}$. Note then that since we've arranged
that for close together venues, they and their metric tensors are
indistinguishable we have $P_{\Theta1}\left(g_{\mu\nu}\right)$= $P_{\Theta2}\left(g_{\mu\nu}\right)$
where $\Theta1$ and $\Theta2$ are two close together venues. $P_{\Theta n}\left(g_{\mu\nu}\right)$
is to be interpreted as the probability of metric$g_{\mu\nu}$ at
venue $\Theta n$.

If one inserts a test particle into the space-time, with small position
and momentum uncertainties, the particle (probability) motion is given
by the Euler-Lagrange equations,

\[
\ddot{x}^{i}+\left\{ _{jk}^{i}\right\} \dot{x}^{j}\dot{x}^{k}=0,
\]
where $\left\{ _{jk}^{i}\right\} $ are the Christoffel symbols of
the second kind, and where $\dot{x}^{j}\equiv dx^{j}/ds$ where \emph{s}
can be either proper time or any single geodesic parameter. Since
$\tilde{g}_{\mu\nu}$ looks stochastic, these equations generate not
a path, but an infinite collection of paths, each with a distinct
probability of occurrence. That is to say that $\left\{ _{jk}^{i}\right\} $
appears stochastic (i.e. $\tilde{\left\{ _{jk}^{i}\right\} }$). Note
that because of the apparent stochasticity of the metric tensor, a
particle initially at rest is unlikely to stay at rest..

In the absence of near by masses, the test particle motion is easily
soluble. Let the particle initially be at (space) venue $\Theta_{0}$.
After time $dt$, the Euler Lagrange equations yield a distribution
of position $D_{1}(x)$ where $D_{1}(x)$ represents the probability
of the particle being in the region bounded by $x$ and $x+dx$. After
another interval $dt$, the resulting distribution is $D_{1+2}(x)$.
From probability theory, this is the convolution,

\[
D_{1+2}(x)=\int_{-\infty}^{\infty}D_{1}(y)D_{1}(x-y)dy.
\]

In this case, $D_{1}(x)=D_{2}(x).$ This is so because, since the
test particle is far from other masses, the Euler-Lagrange equation
will give the same distribution $D_{1}(x)$ regardless of at which
point one propagates the solution. That is to say that $g_{\mu\nu}(x_{1})$,
$g_{\mu\nu}(x_{2})$, $g_{\mu\nu}(x_{3})....$are identically distributed
random matrices. Thus $D_{1}(x),$ $D_{2}(x),$ $D_{3}(x)....$ are
identically distributed random variables. The motion of the test particle
is the repeated convolution $D_{1+2+3....n.....}(x)$, which by the
central limit theorem is a normal distribution. Thus the position
(probability) spread of the test particle at any time $T>0$ is a
Gausssian. The spreading velocity is found to be a constant because
of the following: After $N$ convolutions ($N$ large), one obtains
a normal distribution with variance $\sigma^{2}$ which, again by
the central limit theorem, is $N$ times the variance of $D_{1}(x)$.
Call the variance of $D_{1}(x)$, $a$. (i.e. $var(D_{1})=a$. The
distribution $D_{1}$is obtained after time $dt$. After $N$ convolutions
then, 

\[
\Delta x=Var\left(D_{\sum_{0}^{n}i}\right)=Na.
\]
 This is obtained after $N$ time intervals $dt.$ One then has,

\[
\frac{\Delta x}{\Delta t}=\frac{Na}{N},
\]
 which is to say that the initially localized test particle probability
(i.e, our knowledge of the trajectory) spreads with a constant velocity
$a$. This is an expression of the spread of the wave function for
a free particle.

In the preceding, we've made use of various equations relating to
dynamics. So it might be appropriate to say what equations mean in
an apparently stochastic space-time.

Since in our model our knowledge of the actual venues (points) of
the space-time has a seemingly stochastic nature, these venues cannot
be used as a basis for a coordinate system nor can derivatives be
formed. However, the space-time of common experience (i.e. the laboratory
frame) is non-stochastic in the large. It is only in the micro world
that the apparent stochasticity is manifested. One can then take this
large-scale nonstochastic space-time and mathematically continue it
into the micro region. This mathematical construct provides a nonstochastic
space to which the stochastic/chaotic physical space can be referred.
The (physical) stochastic coordinates $\tilde{x}^{i}$ then are stochastic
only in that the equations transforming from the laboratory coordinates
$x^{i}$ to the physical coordinates $\tilde{x}^{i}$are stochastic.

We'll now use the contravariant observable postulate to derive the
uncertainty relation for position and momentum. Similar arguments
can be used to derive the uncertainty relations for other pairs of
conjugate variables. It will also be shown that there is an isomorphism
between a variable and its conjugate, and covariant and contravariant
tensors.

We assume that we're able to define a Lagrangian, $L.$ One defines
a pair of conjugate variables in the usual way,

\[
p_{j}=\frac{\partial L}{\partial\tilde{q}^{j}}.
\]
 Note that this defines $p_{j}$ a covariant quantity. So that a pair
of conjugate variables so defined contains a covariant and a contravariant
member (e.g. $p_{j}$and $q^{j}$). But since $p_{j}$ is covariant,
it (because of the contravariant observable postulate) is not observable
in the laboratory frame. The observable quantity is just,

\[
\tilde{p}^{j}=\tilde{g}^{j\nu}p_{\nu}.
\]
But $\tilde{g}^{j\nu}$ appears stochastic then so too will be $\tilde{p}^{j}$.
Thus if one member of an observable conjugate variable pair is well
defined, the other member is stochastic. To derive an uncertainty
relation for a conjugate variable pair, consider the following,

\[
\triangle q^{1}\triangle p^{1}=\triangle q^{1}\triangle\left(p_{\nu}\tilde{g}^{\nu1}\right).
\]
What is the minimum value of this product? Since $p$ is an independent
variable, we may take (for the moment) $\triangle p_{j}=0$ so that
\[
\triangle p^{1}=\triangle\left(p_{\nu}\tilde{g}^{\nu1}\right)=p_{\nu}\triangle\tilde{g}^{\nu1}.
\]
In order to determine $\triangle\tilde{g}^{\nu1}$ we will argue that
the variance of the distribution of the average of the metric tensor
over a region of space-time is inversely proportional to the volume,
V, i.e.,
\[
Var\left(\frac{1}{V}\int_{v}\tilde{g}_{\mu\nu}d\mu d\nu\right)=\frac{k}{V}
\]
 In other words, we wish to show that if we are given a volume and
if we consider the average values of the metric components over this
volume, then these average values, which of course appear stochastic,
appear less stochastic than the metric component values at any given
venue in the volume, and that the stochasticity, which we can represent
by the variances of the distributions of the metric components, is
inversely proportional to the volume. This allows that over macroscopic
volumes, the metric tensor behaves classically (i.e. according to
general relativity).

As an idealization, let's assume the distribution of each metric tensor
component at any venue $\Theta$ is Gaussian,

\[
f_{\tilde{g}_{\mu\nu}}(g_{\mu\nu})=\frac{1}{\surd2\pi\sigma}e^{-\frac{1}{2}\left(\frac{g_{\mu\nu}}{\sigma}\right)^{2}}.
\]
 Note also that if $f(y)$ is a Gaussian distribution, the scale transformation
$y\longrightarrow y/m$ results in $f(y/m)$ which is Gaussian with

\[
\sigma_{(y/m)}^{2}=\frac{\sigma_{y}^{2}}{m^{2}}
\]
 Also, we define $f_{g_{\mu\nu}}$at $e_{1}(g_{\mu\nu})\equiv f_{\Theta_{1}}(g_{\mu\nu}).$
We now require

\[
Var(f_{((\Theta_{1}+\Theta_{2}+....+\Theta_{m})/m})\equiv\sigma_{((\Theta_{1}+\Theta_{2}+....+\Theta_{m})/m),}^{2}
\]
where $f_{(\Theta)}$ is normally distributed. Now again, the convolute
$f_{\left(\Theta_{1}+\Theta_{2}\right)}(g_{\mu\nu})$ is the distribution
of the sum of $g_{\mu\nu}$ at $\Theta_{1}$ and $g_{\mu\nu}$ at
$\Theta_{2}$,
\[
f_{\left(\Theta_{1}+\Theta_{2}\right)}=\int_{-\infty}^{\infty}f_{\Theta_{1}}(g_{\mu\nu}^{1})f_{\Theta_{2}}(g_{\mu\nu}^{1}-g_{\mu\nu}^{2})dg_{\mu\nu}^{2},
\]
 where $g_{\mu\nu}^{1}$ is defined to be $g_{\mu\nu}$ at $\Theta_{1}$.
Here $f_{\Theta_{1}}=f_{\Theta_{2}}$ as there are presumed to be
no masses in the neighborhood of the test particle. So that,
\[
f_{(\Theta_{1}+\Theta_{2})/2}=f_{(g_{\mu\nu}/2}at_{\Theta_{1}+g_{\mu\nu}/2}at_{\Theta_{2})}
\]
is the distribution of the average of $g_{\mu\nu}$ at $\Theta_{1}$
and $g_{\mu\nu}$ at $\Theta_{2}.$ $\sigma_{\left(\Theta_{1}+\Theta_{2}+....\Theta_{m}\right)}^{2}$
is easily shown from the theory of normal distributions to be,
\[
\sigma_{\left(\Theta_{1}+\Theta_{2}+....+\Theta_{m}\right)}^{2}=m\sigma_{\Theta}^{2}.
\]
Also, $f_{\left(\Theta_{1}+\Theta_{2}....+\Theta_{m}\right)}$ is
normal. Hence,
\[
\sigma_{((\Theta_{1}+\Theta_{2}+....+\Theta_{m})/m)}^{2}=\frac{m\sigma_{\Theta}^{2}}{m^{2}}=\frac{\sigma_{\Theta}^{2}}{m},
\]
or the variance is inversely proportional to the number of elements
in the average, which in our case is proportional to the volume. For
the case where the distribution $f_{\left(g_{\mu\nu}\right)}$ is
not normal, but also not 'pathological', the central limit theorem
gives the same result as the normal case. Further, if the function
$f_{\left(g_{\mu\nu}\right)}$ is indeed not normal, the distribution
$f_{((\Theta_{1}+\Theta_{2}+....+\Theta_{m})/m)}$ in the limit of
large $m$ is normal,

\[
f_{((\Theta_{1}+\Theta_{2}+....+\Theta_{m})/m)}\longrightarrow f_{((\int_{V}\tilde{g_{\mu\nu}dV)/V).}}
\]
In other words, over any finite region of space-time, the distribution
of the average of the metric tensor over the region is Gaussian. Therefore,
in so far as we do not consider particles to be point sources, we
may take the metric fluctuations around the location of a particle
as normally distributed for for each of the metric components $\tilde{g}_{\mu\nu}.$
Note however that this does not imply that the distributions for any
of the metric tensor components are the same for there is no restriction
on the value of the variances $\sigma^{2}$. Note also that the condition
of normally distributed metric components does not restrict the possible
particle probability distributions, save that they be single-valued
and non-negative. This is equivalent to the easily proved statement
that the functions
\[
f_{\left(x,\alpha,\sigma\right)}=\frac{1}{\surd2\pi\sigma}e^{\left(-\frac{1}{2}\left(\frac{x-\alpha}{\sigma}\right)^{2}\right)}
\]
 are complete for non-negative functions.

Having established that,

\[
Var\left(\frac{\Theta_{1}+\Theta_{2}+....+\Theta_{m}}{m}\right)=\frac{\sigma_{\Theta}^{2}}{m},
\]
 consider again the uncertainty product $\triangle q^{1}\triangle p^{1}=p_{\nu}\triangle q^{1}\triangle\tilde{g}^{\nu1}.$
$\triangle q^{1}$ goes as the volume (volume here is $V^{1}$ the
one-dimensional volume). $\triangle\tilde{g}^{\nu1}$ goes inversely
as the volume, so that $p_{\nu}\triangle q^{1}\triangle\tilde{g}^{\nu1}$
is independent of the volume; i.e. as one takes $q^{1}$ to be more
localized, $p^{1}$ becomes less localized by the same amount, so
that for a given covariant momentum $p_{j}$ (which we might call
the proper momentum), $p_{\nu}\triangle q^{1}\triangle g^{\nu1}=$
a constant k. If $p_{\nu}$ is also uncertain, $p_{\nu}\triangle q^{1}\triangle g^{\nu1}\geq k$
or equivalently,
\[
\triangle q^{1}\triangle p^{1}\geq k
\]
which is the uncertainty principle.

\section{The Wave Function, the Two-slit Experiment, Measurement, and the
Arrow of Time}

If we are to derive the results of quantum mechanics purely from characteristics
of the metric tensor, we need to somehow identify the quantum mechanics
wave function $\Psi$ as some function of the metric tensor. While
we can easily identify the probability density$\Psi^{*}\Psi$ with
$\sqrt{-\left\Vert g_{\mu\nu}\right\Vert }$ it is not immediately
clear how to treat $\Psi$ by itself. The utility of $\Psi$ is that
it contains phase information. Hence using $\Psi$ allows interference
phenomena. One might think then, that our stochastic space-time approach
might have considerable difficulty in producing interference. If however,
we assume a particle does indeed have an associated DeBroglie wavelength
(we'll attempt to explain the genesis of the DeBroglie wavelength
in Part II), then the metric superposition postulate can generate
interference as follows:

Again, consider the free particle in space where there is no nearby
mass. this condition implies that over a region of space near the
particle, the metric tensor is, on average, Minkowskian. And again,
the position probability density $P(x,t)=\sqrt{-\left\Vert g_{\mu\nu}\right\Vert }$.
Now consider a two-slit experiment. Let the situation $s1$ where
only one slit is open result in a metric (where the stochastic elements
are averaged out) $g_{\mu\nu}^{s1}.$ And the case where only slit
two is open, $s2$, result in $g_{\mu\nu}^{s2}$. The case where both
slits are open is then (by postulate 5) is $g_{\mu\nu}^{s3}=\frac{1}{2}\left(g_{\mu\nu}^{s1}+g_{\mu\nu}^{s2}\right)$.
Once we've justified the particle's DeBroglie wavelength, this will
provide for interference. Note that even if the particles are sent
to the screen one-by-one, the metric tensor $g_{\mu\nu}^{s3}$ still
gives the probability density of the particle landing at any position
on the screen and hence still gives interference. (In most of the
remainder of Part I, we assume the metric stochasticity is averaged
out and so we'll omit the stochasticity tilda over $g_{\mu\nu}.$)

With interference and the wave function $\Psi$ in mind, what more
can we say about metric $g_{\mu\nu}^{s1}$? Assume a particle is traveling
in, say, the $x^{3}$direction and, of course, the $x^{4}$ direction.
We might expect the metric, after averaging out the stochasticty to
be the Minkowski metric, $\eta_{\mu\nu}$ save for $g_{33}$ and $g_{44}$,

\[
g_{\mu\nu}^{s1}=\begin{vmatrix}1 & 0 & 0 & 0\\
0 & 1 & 0 & 0\\
0 & 0 & a & 0\\
0 & 0 & 0 & -b
\end{vmatrix},
\]
 where $a$ and $b$ are as yet undefined functions. In order that
the probability density be constant, we need $\left\Vert g_{\mu\nu}^{s1}\right\Vert =-ab$
to be constant. We'll take $a=b^{-1}$ so that $\left\Vert g_{\mu\nu}^{s1}\right\Vert =\left\Vert \eta_{\mu\nu}\right\Vert =-1$.

Now, for the moment, we'll introduce an unphysical situation. Let
$a=e^{i\alpha}$ where $\alpha$ is some as yet unspecified function
of position. Consider the following metrics,

\[
g_{\mu\nu}^{s1}=\begin{vmatrix}1 & 0 & 0 & 0\\
0 & 1 & 0 & 0\\
0 & 0 & e^{i\alpha} & 0\\
0 & 0 & 0 & -e^{-i\alpha}
\end{vmatrix},
\]
\[
g_{\mu\nu}^{s2}=\begin{vmatrix}1 & 0 & 0 & 0\\
0 & 1 & 0 & 0\\
0 & 0 & e^{i\beta} & 0\\
0 & 0 & 0 & -e^{-1\beta}
\end{vmatrix},
\]
where $\alpha$ and $\beta$ are some unspecified functions of position.
For the metrics, $\sqrt{-\left\Vert g_{\mu\nu}^{s1}\right\Vert }=\sqrt{-\left\Vert g_{\mu\nu}^{s2}\right\Vert }=1$,
and noting that for a 4x4 matrix, $\left\Vert \frac{1}{2}A\right\Vert =\frac{1}{16}\left\Vert A\right\Vert $,
\[
\sqrt{-\left\Vert g_{\mu\nu}^{s3}\right\Vert }=\sqrt{\frac{-1}{16}\left\Vert g_{\mu\nu}^{s1}+g_{\mu\nu}^{s2}\right\Vert }=\sqrt{\frac{-1}{16}\left(2+e^{i\left(\alpha-\beta\right)}+e^{-i\left(\alpha-\beta\right)}\right)}=\frac{1}{2}Abs\left(cos\left(\alpha-\beta\right)\right).
\]
This is, of course, the phenomenon of interference. The metrics $g_{\mu\nu}^{s1},g_{\mu\nu}^{s2},$
and $g_{\mu\nu}^{s3}$ represent, for example, the two-slit experiment
previously described. The analogy of the function $e^{i\alpha}$ with
the wave function $\Psi$ is obvious. However, the use of complex
functions in the metric is unphysical as the resultant line element
$ds^{2}=g_{\mu\nu}dx^{\mu}dx^{\nu}$ would be complex. But could we
reproduce the previous scheme, but with real functions? The answer
is yes, but first we must briefly discuss quadratic-form matrix transformations\cite{F9}.
Let,

\[
X=\begin{vmatrix}dx^{1}\\
dx^{2}\\
dx^{3}\\
dx^{4}
\end{vmatrix},
\]
and let matrix $G=g_{\mu\nu}$. Then $X^{t}GX=ds^{2}=g_{\mu\nu}dx^{\mu}dx^{\nu},$
where $X^{t}$ is the transpose of $X.$ Consider transformations
which leave the line element $ds^{2}$ invariant. Given a transformation
matrix $W,$ we can have $X=WX'$ and $X^{t}GX=X'^{t}G'X'=(X^{t}(W^{t})^{-1})G'(W^{-1}X).$
{[}Note: $(WX')^{t}=X'^{t}W'^{t}$.{]} However, $X^{t}GX=(X^{t}(W^{t})^{-1})(W^{t}GW)(W^{-1}X)$
so that $G'=W^{t}GW.$ In other words, the transformation $W$ takes
$G$ into $W^{t}GW.$

Now in the transformed coordinates, a metric $g_{\mu\nu}^{s1}\equiv G^{s1}$
goes to $W^{t}G^{s1}W.$ Therefore $\Psi_{1}^{*}\Psi_{1}=\sqrt{-\left\Vert W^{t}G^{1}W\right\Vert }=\sqrt{-\left\Vert W^{t}\right\Vert \left\Vert G^{1}\right\Vert \left\Vert W\right\Vert }.$
And $\Psi_{3}^{*}\Psi_{3}=\sqrt{-\frac{1}{16}\left\Vert W^{t}\right\Vert \left\Vert G_{1}+G_{2}\right\Vert \left\Vert W\right\Vert }.$

If we can find a transformation matrix $W$ with the properties,

\qquad{}(i) $\left\Vert W\right\Vert =1,$

\qquad{}(ii) $W$ is not a function of $\alpha$ or $\beta,$

\qquad{}(iii) $W^{t}GW$is a matrix with only real components,

then we will again have the interference phenomenon with $g_{\mu\nu}^{'}$
real, $\Psi_{1}^{*}\Psi_{1}=\Psi_{2}^{*}\Psi=1,$ and $\Psi_{3}^{*}\Psi_{3}=\frac{1}{2}Abs\left(cos\frac{\alpha-\beta}{2}\right).$
The appropriate matrix $W$ is,

\[
W=\begin{vmatrix}1 & 0 & 0 & 0\\
0 & 1 & 0 & 0\\
0 & 0 & \frac{-i}{\sqrt{2}} & \frac{1}{\sqrt{2}}\\
0 & 0 & \frac{1}{\sqrt{2}} & \frac{-i}{\sqrt{2}}
\end{vmatrix}.
\]

If, as previously,

\[
G^{s1}=\begin{vmatrix}1 & 0 & 0 & 0\\
0 & 1 & 0 & 0\\
0 & 0 & e^{i\alpha} & 0\\
0 & 0 & 0 & -e^{-i\alpha}
\end{vmatrix},
\]

then,

\[
W^{t}G^{s1}W=\begin{vmatrix}1 & 0 & 0 & 0\\
0 & 1 & 0 & 0\\
0 & 0 & -cos\left(\alpha\right) & sin\left(\alpha\right)\\
0 & 0 & sin\left(\alpha\right) & cos\left(\alpha\right)
\end{vmatrix},
\]
so that in order to reproduce the phenomenon of interference, the
transformed metric tensor will have off-diagonal entirely real terms.
The coordinates appropriate to $G'$ are $x^{1'}=x^{1},$ $x^{2'}=x^{2},$
$x^{3'}=\frac{-i}{\sqrt{2}}x^{3}+\frac{1}{\sqrt{2}}x^{4}$ and $x^{4'}=\frac{1}{\sqrt{2}}x^{3}-\frac{i}{\sqrt{2}}x^{4},$
which is to say that with an appropriate coordinate transformation
(which is complex), we can treat the probability distribution $\Psi^{*}\Psi$
in an intuitive way. In so far as differential geometry is coordinate
independent, we can simply ignore that the coordinate system is complex.

Incidentally, if we look at the sub-matrix,

\[
\begin{vmatrix}-cos\left(\alpha\right) & sin\left(\alpha\right)\\
sin\left(\alpha\right) & cos\left(\alpha\right)
\end{vmatrix},
\]
it is worth noting that this represents a \emph{rotoreflection} transformation,
that is to say a rotation accompanied by a reflection. A repeated
application of this transformation is suggestive of snapshots of a
torsional vibration. We'll have cause to explore torsional vibrations
in Part II.

Although we've explored superposition, we have yet to provide an explanation
for the measurement problem and the 'delayed choice' phenomenon in
the two-slit experiment. First we need to discuss the measurement
process at the slits--and also locality and objective realism.

Consider the two slit experiment using electrons. As an electron 'passes
through' a slit, its electric field must distort the electrons in
atoms at the surface of the slit. As such, it is a measurement of
sorts. But as the electron continues through, the slit electrons of
the slit atoms return to their previous states. So the 'measurement'
is not preserved. The film can be run backward and it would be a valid
physical situation. For there to be a true measurement then, there
must be a mechanism to 'remember' the measurement \cite{key-1}--
a latch or flip-flop of sorts. And that would mean the film could
\emph{not} be run backward. We regard measurement then, as a breaking
of time-reversal symmetry. 

As it is usually maintained, a description of entanglement requires
abandonment of the concepts of objective realism and also locality
(We'll deal with entanglement in Part II). Indeed, Bell's theorem
requires that we must abandon at least one of the two concepts. Dropping
objective reality means that a physical state isn't defined until
it is measured (e.g. is the cat dead or alive?). and dropping locality
means that things separated in space can influence each other instantaneously
(e.g. the collapse of the wave function).

Our model, while preserving objective realism, is non-local. Further,
our non-locality allows for those fluctuations to move backward in
time (except where measurements forbid it), retracing their paths.

Now again consider the two-slit experiment with slits $A$ and $B,$
and a screen at the rear of the experiment. Further, let there be
a detector at $A$ which triggers when a particle goes through slit
$A.$ 

In accord with our model, there are a number of points to be made:

1) A particle will go through only \emph{one} slit. Which one depends
on the stochastic fluctuations of the metric.

2) The metric fluctuation (pilot wave) carries frequency information
(via the determinant of the metric tensor) of the particle.

3) The pilot wave goes through \emph{both} slits.

4) The probability of a particle hitting the screen is again determined
by the determinant of the metric (the differential volume element).

5) In order to explain a measurement at a slit destroying the interference
pattern, we'll posit that the interference phenomena are very fragile,
and any disturbing of the metric fluctuations can wipe out the interference
information. But, if the disturbance is not ongoing, the fluctuations,
as they are able to move freely in time, can go back in time and re-establish
the interference at a place where there was no disturbance.. So if,
for example, the particle goes through slit B, the detector at A will
continue to operate (exerting a field in the vicinity of the slit)
and the interference cannot re-establish. In the case where the particle
goes through slit A, once detected, the detector can be switched off.
But in this case the interference cannot re-establish since the disturbance
cannot propagate backwards through a measurement (a flip-flop).

6) As the model has fluctuations being propagated backward in time,
the delayed choice experiment follows the same arguments as the above.

Our non-locality then, requires access to the past (at least for small
metric fluctuations) and so raises questions as to the arrow of time.
The arrow of time' seems to emerge from statistical mechanics (via
entropy). And statistical mechanics differs from mechanics in that
there are many particles in play, and the particles interact. So it
may well be that the arrow of times is a result of particle interactions.

For an isolated particle though, there's maybe no arrow of time, or
more likely a very small arrow resultant from the slight time-reversal
symmetry breaking in the weak interactions. Indeed, if there were
no slight bias for an arrow of time, the universe as a whole wouldn't
display one. When a large ensemble of particles interact, the arrow
likely grows longer. (Whilst a particle can be run backward in time,
breaking an egg can't.) Further, in the macro-world, everything is
a measurement of sorts (viewing a scene gives an estimate of positions,
etc.) and hence we can't run macro-world scenes backwards; a strong
arrow of time has been established. 

While the model is a mechanical description, it is based on an underlying
stochasticity of spaced-time. So it seems that God does indeed play
dice.

\part{Crypto-stochastic Space-time}

There are (at least) two problems with the stochastic space-time model:
First, there's just so far one can take stochasticity. Almost by definition,
it is difficult to derive deterministic equations from stochastic
elements. The second and more serious problem is that our model posits
a stochastic metric tensor while requiring that the determinant of
the metric be non-stochastic. the square root of minus the determinant
is identified with a well-defined probability density. While mathematically
it is easy to get a non-stochastic determinant from a matrix with
stochastic elements, it's difficult to justify with physics.

But we don't actually require stochasticity in the metric elements;
we just require that they \emph{appear} stochastic--in the sense that
repeated measurements give unpredictable results. One way of obtaining
this is to replace the stochasticity with an immeasurably high frequency
fluctuation in the elements. The idea is that the stochastic energy
fluctuations in the vacuum drive the space-time into a collective
oscillatory mode (a kind of stochastic resonance). This assumption
will allow us to illuminate polarization phenomena and even entanglement.
Quantum mechanics then, with this modification, is deterministic but
with aspects that are unmeasurable to arbitrary accuracy. Determinant
but not measurable (along with non-linearity) is the book characterization
of chaos. We'll make use of this in Part III.

The question is: what is the nature of these oscillations. First we'll
see what oscillator models can best elucidate troublesome quantum
phenomena, e.g. entanglement and optical polarization. And our descriptions
must preserve objective reality (a particularly difficult problem
with polarization and entanglement).

We require the metric oscillations to be of a very high frequency,
sufficiently high that we can't measure them. It seems reasonable
to restrict the frequency to below $10^{43}hz$ (which is the frequency
where the wavelength is the Planck length) and above $10^{30}hz$
(the frequency of the highest indirectly measured gamma rays).

Taking as a hint, the roto-reflections mentioned earlier, we'll posit
that the oscillations are torsional around particles (including photons).

\section{Linear Polarization}

Consider now optical linear polarization. We'll address three issues:
1-the reason half the incident photons go through a polarizer, rather
than just photons with polarization oriented in the same direction
as the polarizer's polarization angle (quantum mechanics has a tortuous
explanation); 2-Malus's Law; and 3-the situation when a third polarizer
is inserted between a pair of crossed polarizers.

\begin{figure}
\includegraphics{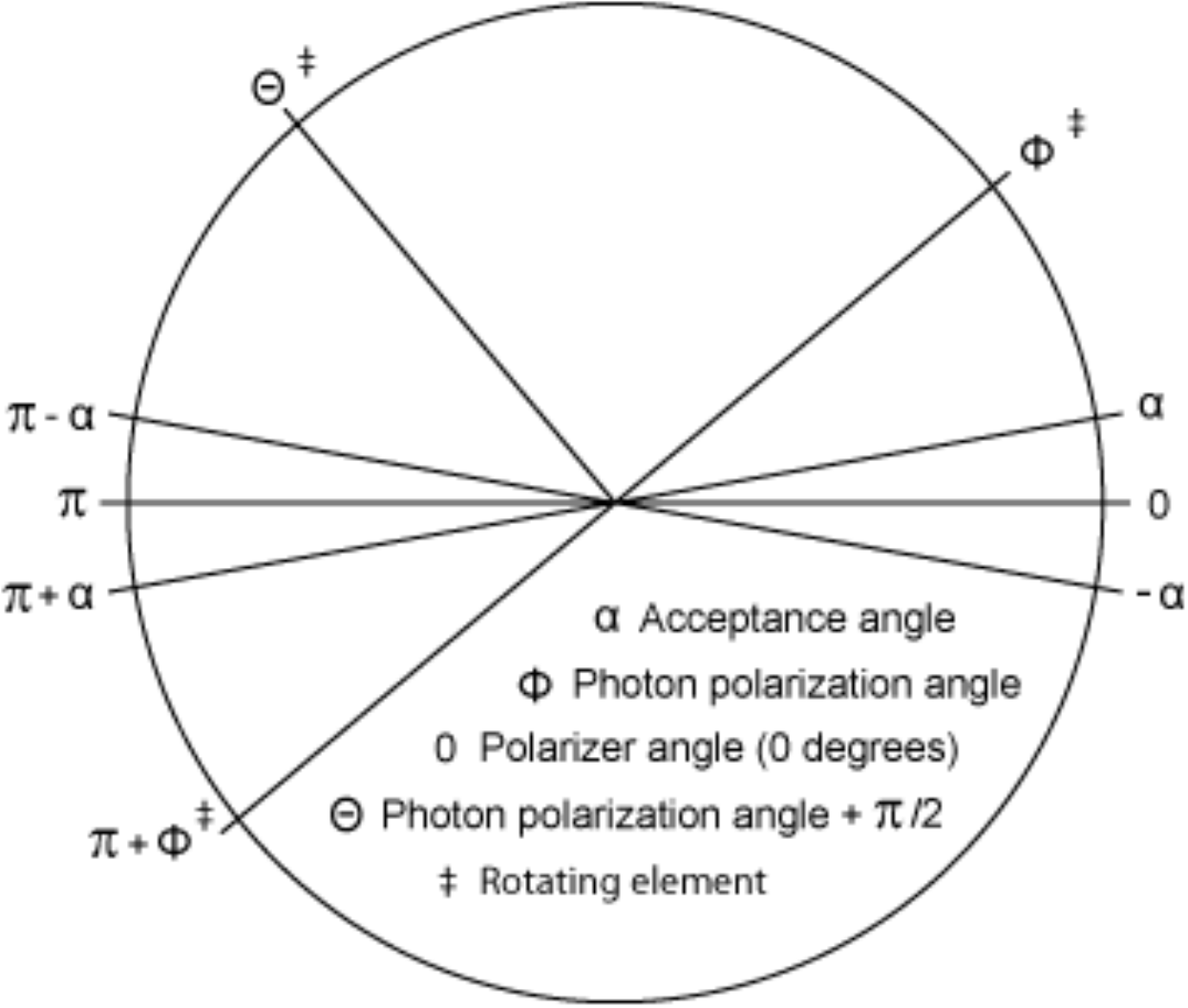}

Fig. 4. A model for optical polarization.
\end{figure}

Consider Figure 4. Assume a polarizer (in the \emph{x}, \emph{y} coordinates
perpendicular to \emph{z}, the direction the photon moves) with polarization
orientation along the $0-\pi$ axis. And consider a photon with polarization
angle $\Phi$encountering the polarizer. Our torsionally oscillating
space-time model assumes the photon is essentially oscillating through
$\pi$ radians around a point $\pi/2$ radians from the polarization
angle. So in figure 4, the photon is oscillating around $\Theta$
from $\Phi$ to $\pi+\Phi.$ (As a short-hand, rather than speaking
of the space-time torsionally oscillating and carrying the photon
with it, we'll refer to it simply as the oscillating photon.) 

A frictionless torsional spring is governed by the equation, $\theta=k*cos(\omega t+\Phi)$
where $\omega$ is the oscillation frequency and $k$ is a constant
(for the moment) involving the torsional spring stiffness and the
angle through which the spring oscillates. At $t=0,$ $\theta$ is
at the extremum, $\Phi.$ 

A photon with a well-defined polarization direction (here $\Phi$)
oscillates as it reaches the polarizer. It is clear that the more
time the rotating photon's polarization vector lies within the acceptance
angles, the more likely the photon will pass through the polarizer.
And so the angular velocity is proportional to the likelihood of the
photon \emph{not} getting through (the attenuation $A$). So \emph{$A_{\theta}$}
(at $\theta)$ is proportional to the angular velocity, $d\theta/dt=-k\omega*sin(\omega t+\Phi).$
Or referenced to $\varphi$ (which is $\theta+\pi/2$) $d\varphi/dt=k\omega*cos(\omega t+\Phi+\pi/2).$
Or $A_{\theta}=k_{1}d\varphi/dt$ where $k_{1}$ is a constant.

The maximum throughput is at $\Phi$ which is at $\pi/2$ from $\Theta,$
and $\theta$ is always equal to $\varphi+\pi/2.$ So as the attenuation
follows a cosine law, so to does the intensity, $I$. We see that
the transmission increases as the cosine of the angle between the
photon polarization angle and the center of the acceptance angle.
I.e. the intensity $I_{\Theta}$at angle $\Theta$ is, $I_{\Theta}=k_{2}d\varphi/dt$
where $k_{2}$ is a constant.

There are three obvious problems: First the transmission amplitudes
are small. With a perfect polarizer; the acceptance angle tends to
a delta function and the transmission become infinitesimal. Second,
the minimum transmission isn't equal to zero. And third, the functional
form of the intensity is wrong; it should go not as cosine, but as
cosine squared. These problems can be handled by considering not only
oscillations in \emph{x} and \emph{y}, but also in \emph{z} and \emph{t}.
Relativity ideas suggest oscillations in all coordinates.

Consider oscillations in the two directions perpendicular to the coordinates
of figure 4, namely $t$ and $z.$ The photons travel along $z$ en
route to the polarizer. If there are oscillations in $t$ against
$z$, the world-line of a photon is not the light cone, a 45 degree
line in a Minkowski diagram, but a wavy line as in figure 5. So, at
any time $t$, the photon exists at a linear series of values of $z$
giving the photon some of the attributes of a 'string' (or dotted
line) of well-defined length.

We assume that the $t,$$z$ oscillation is synchronized with the
$x,$$y$ oscillation and that the zero point is $\pi/2$ displaced
from the axis of rotation just as in the $x$$y$ case. Now consider
a photon with, for example, polarization angle=0 encountering a polarizer
with polarization angle also equal to zero. If when it reaches the
polarizer, the oscillating photon's rotation lines up with the polarizer's
axis (very rarely), it goes through. If not, then the photon is displaced
slightly back in time and with a slightly lower angle. Again, it goes
through, or not. If not, again the photon slightly back in tame and
angle encounters the polarizer. The process continues until either
the photon goes through the polarizer or the time oscillation goes
forward again. So the transmission probability is 1/2, which is what
it should be. Now, using the same argument as with the $x,$$y$ case,
if the photon polarization is at an angle $\phi$ with respect to
the polarizer angle, then the probability of transmission due to the
oscillations goes as the cosine of the angle. So considering both
time and space oscillations gives an attenuation of cosine squared
of the angle. And that is the expected result, i.e. Malus's law. When
the photon enters the polarizer, it is 'prepared' by the polarizer.
That is to say that since the polarizer can admit only photons with
polarization direction the same as that of the polarizer, the photon
is forced to the polarization of the polarizer. The photon continues
to rotate, but (usually) around a different angle.

In the case of two crossed polarizers, there is nothing new; no light
gets through. But if a third polarizer at, say, a 45 degree angle
is interposed between the two crossed polarizers, 1/4 of the light
gets through. The conventional quantum mechanics explanation is that
when the photon encounters the interposed polarizer, it decomposes
into components parallel and perpendicular to the polarization angle
of the interposed polarizer, and then either gets transmitted or absorbed
with a probability based on the amplitudes of the decomposed polarization
vector.

Our model explains this effect as follows: when the oscillating photon
encounters the interposed polarizer, it, as described earlier, goes
through with some probability. But as it does so, it is 'prepared'
(as described above) to have the same polarization angle as the interposed
polarizer. So now the newly prepared photon has a polarization angle
that is no longer at a right angle to the third polarizer. So there
is a probability of the photon getting through the third polarizer. 

Using the word 'probability' might seem to imply that the transmission
is, in the quantum mechanical sense, probabilistic. But it is actually
deterministic in principle. If we knew the rotational angle of the
photon when it encountered the polarizer, we would know if the photon
would or would not get through. In practice though, since the rotational
frequency is way too high to measure, in practice, all we can give
is a probability.

\begin{figure}
\includegraphics{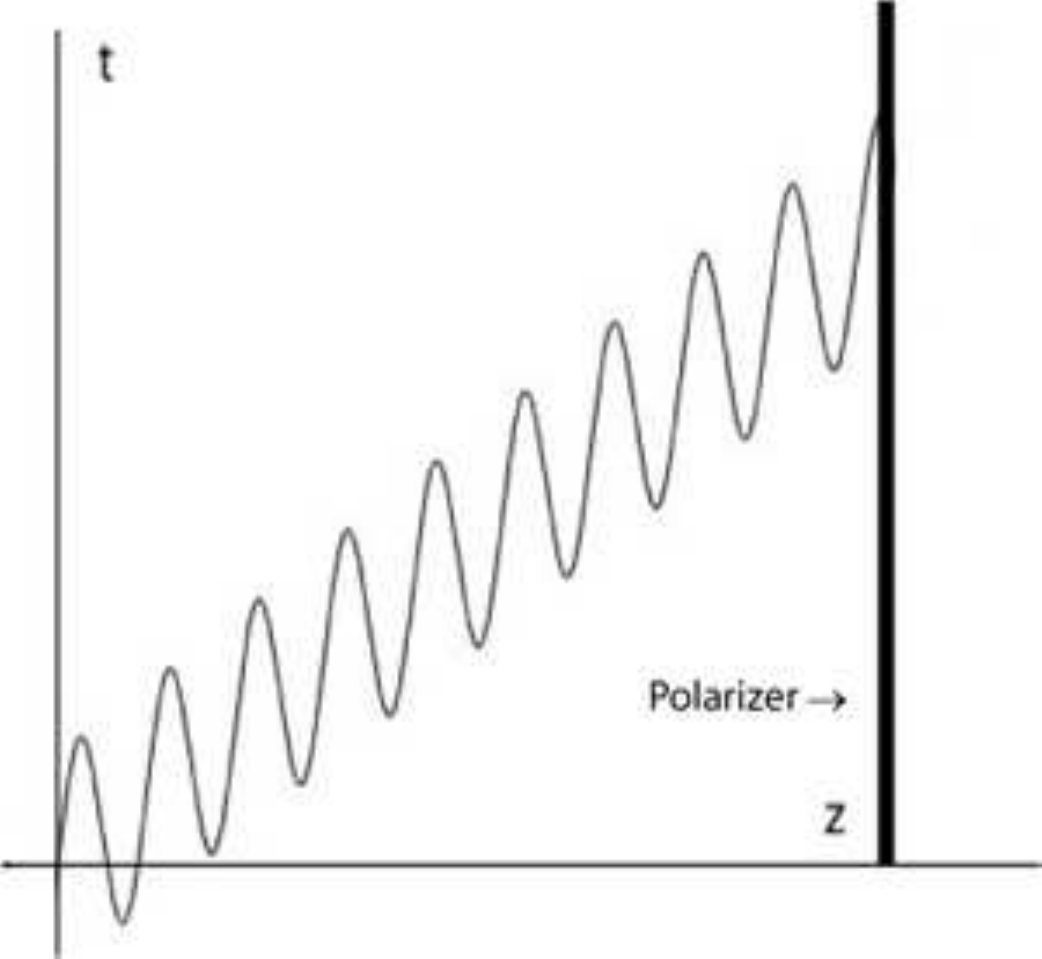}

Fig 5. Oscillations of $t$ against $z$ 

as photon approaches polarizer.
\end{figure}

\section{A Note on Entanglement}

As the model seems to give a model of polarization, one might ask
if the model has anything to say about entanglement. It might. Usually,
a pair of entangled particles emerge from a single venue. One might
have that (in our model) the particles are locked to exactly the same
oscillation behavior (or perhaps are locked $\pi$ radians out of
phase). And when they separate, they remain so locked. In the language
of quantum mechanics, they share a common wave function. When one
of the particles is measured, its oscillation freezes at its current
oscillation angle. Again, in the language of quantum mechanics, the
wave function collapses. In our language, the metric distortion dissipates
and the other particle also freezes at the same angle (or the same
angle $+\pi$). This model also addresses single particle spin up/down
measurements in Stern-Gerlach experiments (where the angle of the
measurement apparatus is varied). Note that this model is, of course,
non-local, as required by Bell's theorem \cite{key-9}. And finally,
as the model incorporates objective reality, Schrödinger's cat is
either alive or dead but not both. 

In this crypto-stochastic model, after stochastic resonance creates
the oscillations, the model is completely deterministic. So perhaps
God does play dice, but only to set things going.

\part{Chaotic Space-time (soon to come)}

The introduction of stochastic space-time admits a phenomenological
explanation of some elements of quantum mechanics. The extension to
crypto-stochastic space-time yields possible explanations of more
quantum phenomena. The aim is to provide a mechanical system to model
all the elements of quantum mechanics. There is much left to do. The
principle task is to explain the origins of the space-time oscillations,
and to provide 'field equations' as in general relativity, to give
quantitative results (from which, the Schrödinger equation, among
others, will drop out). 

The first thing to note is that the present model, derived as it is
from ideas of General Relativity, is inherently non-linear, although
as the constant of gravitation is very small (compared to the electro-weak
force) the non-linear effects are exceedingly small. Further, the
model is (once the oscillations are established) completely deterministic.
Further, though deterministic, the state (phase angles) of the oscillations
is unmeasurable. But these are the defining elements of a chaotic
system: non-linearity, deterministic, immeasurable. We will demonstrate
(in a forthcoming paper) that do to stochastic resonance, the stochastic
space-time has large-scale periodicity. And this deterministic periodicity
allows chaos which then allows small-scale, self-organizing periodicity
at the scale of the elementary particles. We use the techniques of
chaos theory to obtain, if not the quantum mechanics analogy of the
relativity field equations, the behavior of space-time at small dimensions.
It should be noted that nonlinear versions of the Schrödinger, Dirac,
and Klien-Gordon equations exist \cite{G1} and they each exhibit
non-local solutions. What is unknown at the moment is how to reconsile
the nonlocality with special relativity \cite{G2}.

Part of the motivation for going to a chaos description is the reasonable
objection to the model that it does not address the apparently random
nature of radioactive decay times. First one might note that a deterministic
system can exhibit (apparently) random behavior, e.g. the state of
an individual molecule in a gas in thermal equilibrium. But more to
the point: If a radioactive atom is in a given state at a given space-time
venue decays after an interval $\Delta t$, then another atom in a
state and position arbitrarily close to the first might be expected
to decay after about the same $\Delta t.$ But for a chaotic system
that is not the case; close states in phase space genereally evolve
to be not close points. So insofar as the decay time $\Delta t$ is
dependent on the state of the radioactive atom, $\Delta t$ is unpredictable.

We have taken as a jumping off point the current wisdom that there
is a stochastic energy fluctuatiton in the vacuum, and from that we
generated some of the phenomena of quantum mechanics. We could just
as well have posited that the energy fluctuations are not stochastic
but chaotic, thus removing any indeterminancy from the model. But
we wished not to make too sudden a break with current notions of the
vacuum. 

And so, perhaps Einstein was right after all; God does not play dice--or
at the most, exceedingly rarely.
\begin{acknowledgments}
I should like to thank Drs. Norman Witriol and Hans Fleischmann, as
well as Nick Taylor for helpful conversations and ideas.\end{acknowledgments}

\end{document}